\documentclass[reprint,aps,prl,longbiography,twocolumn,10pt,superscriptaddress,nobibnotes]%
			  {revtex4-2}  
\usepackage{graphicx}	
\usepackage{subfigure}
\usepackage{xr} 
\usepackage{dcolumn}  					
\usepackage{bm} 
\usepackage{hyperref}
\hypersetup{colorlinks=true, citecolor=blue, urlcolor=blue, linkcolor=blue}
\usepackage{xcolor}    
\usepackage{soul}      

\usepackage{placeins}

\begin{document}

\title{Interspecies clock comparison below $5 \times 10^{-18}$ uncertainty with a transportable clock}

\author{C. Vishwakarma}
\author{I. Nosske}
\author{T. Lücke}
\author{M. Steinel}
\author{M. Filzinger}
\author{E. Benkler}
\author{S. Dörscher}
\author{N. Huntemann}
\author{C. Lisdat}
\email{christian.lisdat@ptb.de} 
\affiliation{Physikalisch-Technische Bundesanstalt, Bundesallee 100, 38116 Braunschweig, Germany}

\begin{abstract}
We report a measurement of the optical frequency ratio between the \mbox{$^2\mathrm{S}_{1/2}(F=0)$--${^2\mathrm{F}_{7/2}(F=3)}$} electric-octupole (E3) transition of $^{171}$Yb$^{+}$ and the \mbox{$^1\mathrm{S}_0$--${^3\mathrm{P}_0}$} transition of $^{87}$Sr, \mbox{$\nu_{\mathrm{Yb}^{+}}/\nu_\mathrm{Sr} = \mbox{1.495\,991\,618\,544\,900\,588\,1(65)}$}.
Reaching a fractional uncertainty of $4.3 \times 10^{-18}$, this result improves upon the previous best by more than a factor of three and is among the few that meet the requirements for interspecies clock comparisons specified by the roadmap towards the redefinition of the SI second.
The comparison is between a transportable optical lattice clock and a stationary single-ion clock.
It spans a period of nearly two years, during which the transportable clock was intermittently operated off-campus.
The ratio was reproducibly measured during four separate campaigns, which are consistent within their statistical uncertainties.
The results demonstrate reproducible $10^{-18}$ level operation of the transportable clock and thus validate its application for chronometric geodesy and as a transfer standard for inter-institute clock comparisons, e.g., in the absence of optical fiber links.
\end{abstract}
\maketitle

Improved understanding of light-matter interaction and precise control of external perturbations have enabled the development of optical clocks with estimated fractional frequency uncertainties below $10^{-18}$ \cite{aep24,arn24,mar25,lin25c, zha26, jia26b, fil26}. 
Yet, validating clock uncertainty budgets requires direct frequency comparisons between independent systems, preferably among clocks developed and operated by different institutes.
Therefore, the roadmap for the redefinition of the SI second calls for accurate frequency comparisons among clocks operating on both the same and different atomic transitions \cite{dim24}, across one or multiple atomic species.
To date, only a few measurements below the required fractional uncertainty of $5 \times 10^{-18}$ have been reported for interspecies clock comparisons \cite{hau25, aep26, fil26}.

While co-located clock comparisons are straightforward, high-accuracy comparisons among distant institutes rely on specialized optical fiber links \cite{lis16, aka20, sch22a}, limiting uncertainties at the level of about $3 \times 10^{-18}$ due to differential relativistic redshifts \cite{den17, rie20}.
Transportable clocks offer an alternative for linking distant clocks by acting as transfer frequency standards \cite{dim24, icon24} that enable direct on-site comparisons, circumventing these limitations.
Together with fiber links, transportable clocks also allow direct determination of relativistic redshifts \cite{gro18a, icon24}.
Both applications rely crucially on the frequency reproducibility of transportable clocks.
However, reproducing the frequency at the level of $5 \times 10^{-18}$ remains challenging even for state-of-the-art systems permanently housed under well-controlled laboratory conditions.
Demonstrating reproducible, high-accuracy operation after repeated transportation therefore represents a central milestone for transportable clocks. 

In this Letter, we report high-accuracy frequency comparisons between the stationary clock Yb1 \cite{hun16, san19} and the transportable clock Sr4 \cite{nos25} at PTB.
We determine the optical frequency ratio $R = \nu_{\mathrm{Yb}^{+}} / \nu_\mathrm{Sr}$ of the \mbox{$^2\mathrm{S}_{1/2}(F=0)$--${^2\mathrm{F}_{7/2}(F=3)}$} electric-octupole (E3) transition at 642~THz in $^{171}$Yb$^{+}$ and the \mbox{$^1\mathrm{S}_0$--${^3\mathrm{P}_0}$} transition at 429~THz in $^{87}$Sr with a fractional uncertainty of $4.3 \times 10^{-18}$.
The measurement improves upon the previous best determination of this ratio \cite{piz26} by more than a factor of three and constitutes the most accurate interspecies frequency comparison involving a transportable clock to date.
Repeated comparisons of the clock across transportation and operational breaks demonstrate the frequency reproducibility of the transportable clock.
Beyond validating the performance of Sr4, the result provides an accurate reference for other optical clocks.
As the two clock transitions are widely used among optical frequency standards, these results are particularly relevant for ongoing efforts towards the redefinition of the second \cite{dim24}.

\begin{figure*}[t]
    \centering
    \includegraphics[width=1\linewidth]{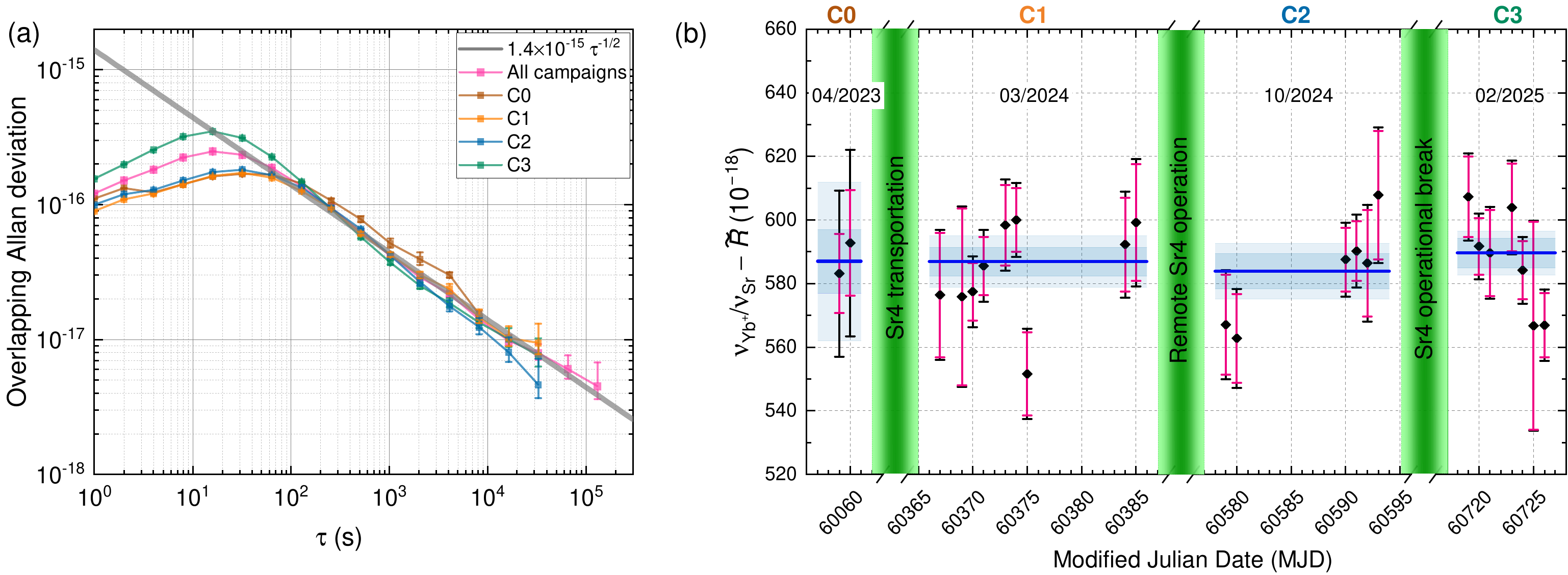}
    \caption{
        Results of the comparisons between the Yb1E3 single-ion clock and the Sr4 transportable optical lattice clock at PTB.
        \textbf{(a)}
        Overlapping Allan deviations of the 
        reduced frequency ratio for each campaign (C0 through C3) and for the combined data. The gray solid line indicates the best fit to the combined data at $\tau \geq 256 \, \mathrm{s}$. 
        \textbf{(b)}
        Daily mean frequency ratios relative to $\widetilde{R} = 1.495 \, 991 \, 618 \, 544 \, 900$.
        The bottom axis shows the Modified Julian Date (MJD), and main measurement months are indicated at the top.
        Magenta (black) error bars indicate the statistical (total) uncertainties. 
        The solid blue lines, dark shaded regions, and light shaded regions represent the mean values, statistical uncertainties, and total uncertainties, respectively, for each campaign.
        The reduced chi-squared value of the daily means across all measurement campaigns with respect to their statistical uncertainties is $\chi^{2}_\mathrm{red} = 1.19$.
    }
    \label{fig:Daily local ratio}
\end{figure*}

The transportable optical lattice clock, Sr4, is housed in an air-conditioned car trailer and has been described in detail previously \cite{nos25}.
$^{87}$Sr atoms are confined in a one-dimensional optical lattice and interrogated in a blackbody radiation shield at a temperature of about $-50 \, ^{\circ}\mathrm{C}$.
Since 2023, the clock's systematic uncertainty has been reduced by nearly an order of magnitude to \mbox{$2.1 \times 10^{-18}$}, largely due to extended characterizations of the lattice light shift and density shift.
We reliably reproduce the electric-dipole (E1) magic frequency in our apparatus, obtaining an overall value of $\nu^{E1} = 368 \, 554 \, 463.4(1.7) \, \mathrm{MHz}$, which is in agreement with the value reported in Ref. \cite{ush18} for a similar configuration.
Therefore, we update the lattice light shift evaluation with respect to our earlier work \cite{nos25} by using the known detuning of the lattice laser from $\nu^{E1}$, rather than campaign-specific evaluations of the detuning based on interleaved measurements at different lattice trap depths.
This approach reduces the need to reevaluate the detuning and lowers the statistical uncertainty thereof, see the Supplemental Material \cite{SupplementalMaterial} for more details.
The frequency instability of Sr4 is around \mbox{$5 \times 10^{-16}/ \sqrt{\tau/\mathrm{s}}$}, where $\tau$ is the averaging time \cite{nos25}.

The stationary ion clock Yb1 uses a single $^{171}$Yb$^{+}$ ion trapped in a radio-frequency (RF) Paul trap. 
A detailed description of this laboratory clock can be found in Refs. \cite{hun16, san19}. 
For the measurements reported in this work, the clock was operated on the E3 transition with fractional frequency uncertainty of $2.7 \times 10^{-18}$. 
The clock is referred to as Yb1E3 to indicate which transition is used. 
The clock laser addressing the reference transition is pre-stabilized to the ultrastable cryogenic silicon cavity Si2 \cite{mat17a}, enabling a frequency instability of about $1 \times 10^{-15}/ \sqrt{\tau/\mathrm{s}}$ for the clock \cite{san19}.

The frequency ratio of the two clocks is determined by comparing, for each system, light from the respective clock laser and from a common, independent ultrastable laser \cite{mat17a} with the output of a single branch of an optical frequency comb.
The two frequency ratios between each clock laser and the ultrastable laser are then combined phase-coherently using the ultrastable laser as a common reference.
This approach suppresses excess noise and systematic shifts between different output ports of the same frequency comb, or even between different frequency combs \cite{ben19_without_erratum}.

The relativistic redshift correction between the two clocks is determined by geometric leveling with an uncertainty of $5.0 \times 10^{-19}$.
While Yb1E3 remains stationary, the height of Sr4 may change after every transport and is thus measured anew with respect to its local reference height marker.

\begin{table*}[tb]
	\caption{\label{tab:freq}
        Summary of the four clock comparison campaigns and the overall average result:
        campaign intervals as MJD; cumulative measurement duration ($T_\mathrm{meas}$); fractional uncertainty contributions from statistics ($u_A$), systematics of the Yb1E3 clock ($u_{B,{\rm Yb1E3}}$), systematics of the Sr4 clock ($u_{B,{\rm Sr4}}$), and the relativistic redshift ($u_{\rm grav}$); fractional combined uncertainty ($u_\mathrm{tot}$); values of the frequency ratio \mbox{$R - \widetilde{R}$}, where $\widetilde{R} = 1.495 \, 991 \, 618 \, 544 \, 900$; and weights $w_i$ in the weighted average.
    }
	\begin{ruledtabular}
	\begin{tabular}{llllllllll}
    Campaign	& MJD	& $T_\mathrm{meas}$ & $u_{A}$ & $u_{B,{\rm Yb1E3}}$& $u_{B,{\rm Sr4}}$& $u_{\rm grav}$& $u_\mathrm{tot}$ &  $R - \widetilde{R}$ & $w_i$\\
		&	& (ks)	&	($10^{-18}$)	& ($10^{-18}$) & ($10^{-18}$)  &  ($10^{-18}$) & ($10^{-18}$) & ($10^{-18}$)\\	
		\colrule
\mbox{C0}	& 60059 -- 60060 &	62.0    & 6.7 &	2.7	&	15	&	0.5	&	16.7  &	587.0(24.9) & $\ll 0.01$ \\
\mbox{C1} 	& 60367 -- 60385 &	222.1	& 3.0 &	2.7	&	3.6	&	0.5	&	5.4  &	586.9(8.1)  & $0.22$ \\
\mbox{C2} 	& 60579 -- 60593 &	155.7 	& 3.7 &	2.7	&	3.5	&	0.5	&	5.8  &  583.9(8.7)  & $0.16$ \\
\mbox{C3} 	& 60719 -- 60726 &	246.5 	& 3.1 &	2.7	&	2.1	&	0.5	&	4.6   &  589.6(6.9) &  $0.63$ \\
\colrule
\mbox{Overall}	&            &	686.3 	& 2.1 &	2.7	&	2.6	&	0.5	&	4.3   &  588.1(6.5) & \\
	\end{tabular}
	\end{ruledtabular}
\end{table*}

The overall measurement consists of four campaigns (C0--C3), which are several months apart. 
During the breaks after campaigns C0 and C1, Sr4 was transported for an inspection of the car trailer and for a remote measurement campaign, respectively. During the break after campaign C2 a faulty laser for the optical lattice was replaced.
Yb1E3 was operational without long breaks.
Figure~\ref{fig:Daily local ratio}~(a) shows the overlapping Allan deviation of the reduced frequency ratio data for each campaign and for the concatenated data of campaigns C0 through C3.
The latter data set exhibits a frequency instability of \mbox{$1.4 \times 10^{-15}/ \sqrt{\tau/\mathrm{s}}$}.
For parts of the measurement, we use frequency stability transfer \cite{hag13} from the ultrastable cryogenic silicon cavity Si2 \cite{mat17a} to the transportable clock laser \cite{her22} of Sr4. While this improves the frequency instability of the local oscillator, it has a negligible effect on the clock comparison instability at long averaging times ($\tau > 10^3~\mathrm{s}$), which is essentially limited by the single-ion clock.

Figure~\ref{fig:Daily local ratio}~(b) shows the daily mean frequency ratios together with the mean values and uncertainties of all four campaigns.
Here, the statistical uncertainties are estimated from the overlapping Allan deviation of the daily data since white frequency noise dominates at long averaging times.
The total uncertainties additionally include the systematic uncertainties of the clocks and of the relativistic redshift corrections.
The results of the four campaigns are summarized in Table \ref{tab:freq}.

We derive the overall average frequency ratio as a weighted average, $\overline{R} = \sum_{i=0}^{3} w_i R_i$, of the frequency ratios $R_i$ measured in the four campaigns, where $\sum_{i=0}^3 w_i = 1$.
The uncertainty of the average value is given by \mbox{$u(\overline{R}) = (\sum_{i,j=0}^{3} w_i C_{ij} w_j)^{1/2}$} where the covariance matrix $C_{ij}$ is derived using the same formalism used in Ref.\ \cite{nos25}, i.e., the covariances introduced by systematic effects are treated as fully correlated between campaigns.
The weights $w_i$ were determined by minimizing $u(\overline{R})$ using the estimated covariance matrix $C_{ij}$ to achieve the best possible estimate.
Notably, this procedure assigns the highest weight to campaign C3 and a negligible weight to campaign C0, see Table \ref{tab:freq}.
The correlation coefficients are reported in the Supplemental Material \cite{SupplementalMaterial}.
We find an overall average frequency ratio $\overline{R} = 1.495\,991\,618\,544\,900\,588\,1(65)$.

The mean frequency ratios per campaign (see Table~\ref{tab:freq}) differ by no more than $3 \times 10^{-18}$ in fractional units from the overall result, which is comparable to the statistical uncertainties of the individual campaigns and highlights the reproducibility of the two clocks' frequencies at this level.
The reduced chi-squared value of the daily [campaign] mean frequency ratios across all four measurement campaigns C0--C3 with respect to their statistical uncertainties is $\chi^{2}_\mathrm{red} = 1.19$ [$0.26$], with a $p$-value of $0.24$ [$0.86$], which is compatible with the estimated uncertainties.

\begin{figure}[tb]
    \centering
    \includegraphics[width=1\linewidth]{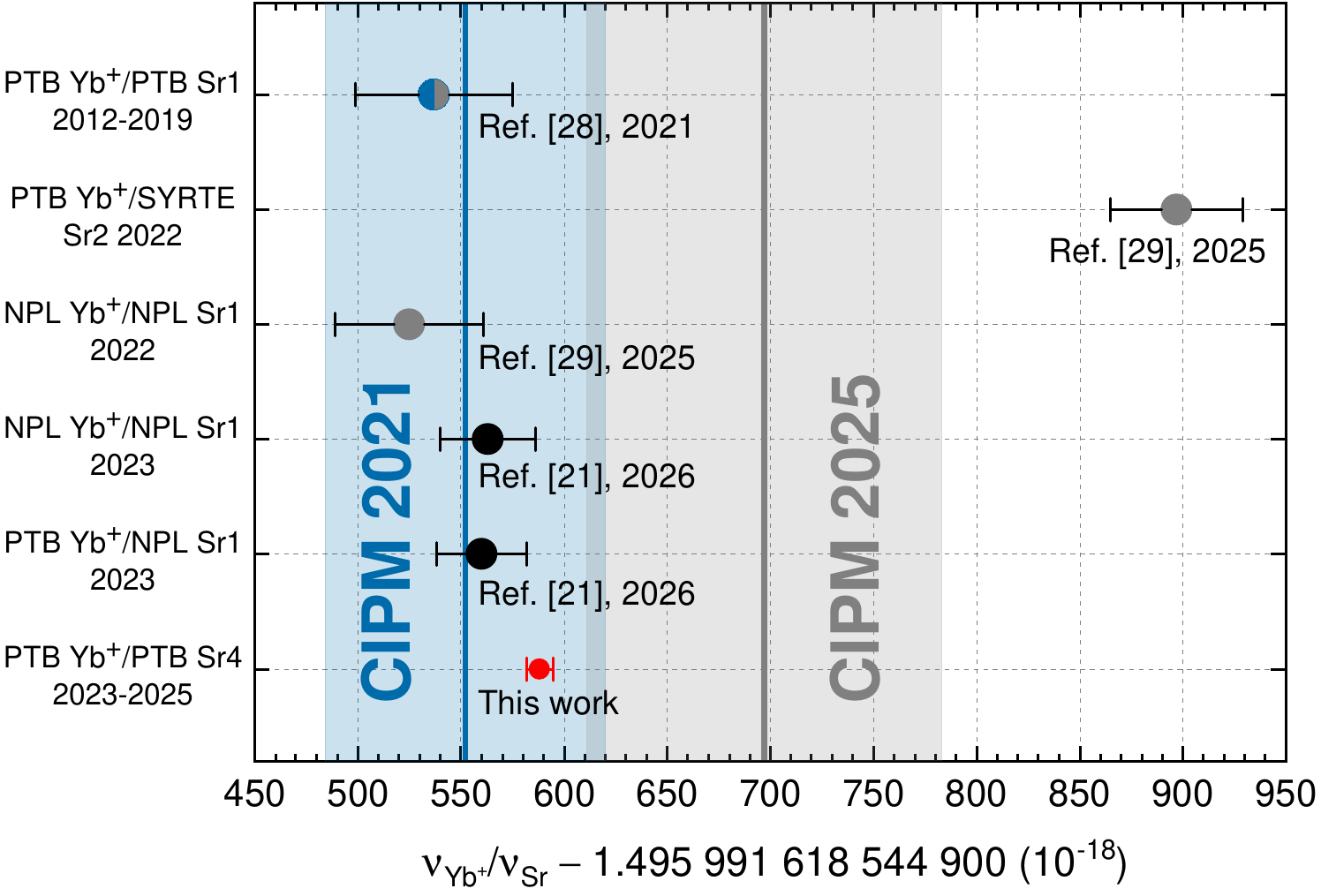}
    \caption{
        Overview of measured values of $\nu_{\mathrm{Yb}^{+}}/\nu_\mathrm{Sr}$ with relative uncertainties below $10^{-16}$ \cite{doe21, lin25a, piz26}.
        On the vertical axis, measurement years are shown, while the labels next to measurement points indicate reporting years.
        The blue and gray lines denote the ratios from the 2021 \cite{mar24a} and 2025 \cite{cipm25} CIPM evaluations, respectively; the corresponding shaded bands indicate their associated uncertainties. The color of each data point reported before 2026 indicates the CIPM evaluation to which it contributed.
        Previous measurements involving the clock PTB-Sr3 are subject to uncontrolled systematic shifts \cite{lin25a, hau25} and thus not shown here.
    }
    \label{fig:Reported ratios}
\end{figure}

We compare our result to previously reported values of the frequency ratio with fractional uncertainties below $10^{-16}$ \cite{ doe21, mar24a, lin25a, cipm25, piz26} in Fig.~\ref{fig:Reported ratios}.
The measured ratio is largely consistent with the measurements at NPL \cite{lin25a, piz26} and PTB \cite{doe21}.
Improving upon the previous best measurement uncertainty \cite{piz26} by a factor of 3.5, it is one of few \cite{hau25, aep26, fil26} to reach the fractional uncertainty level of \mbox{$< 5 \times 10^{-18}$} required by the roadmap for the redefinition of the second \cite{dim24} and the first for this particular frequency ratio.
We note that while the other high-accuracy comparisons were performed between stationary optical clocks in well-controlled laboratory environments, we compare a stationary system to a transportable clock, which has actually been transported between measurements.

Our measurement of the ${}^{171}\mathrm{Yb}^+ (\mathrm{E3}) / {}^{87}\mathrm{Sr}$ frequency ratio further provides a valuable benchmark for investigating and resolving inconsistencies that have been observed in several other measurements \cite{hau25, lin25a, piz26, icon24}, which cannot be explained by a single point of error.
In particular, our result differs significantly from the 2022 comparison between the SYRTE-Sr2 and PTB-Yb1E3 clocks \cite{lin25a}, supporting the conclusion drawn in Ref.~\cite{lin25a} that the SYRTE-Sr2 clock was subject to an uncontrolled frequency error at the $10^{-16}$ level. We note that this potential error propagates into the 2025 adjustment of standard frequencies recommended by the International Committee for Weights and Measures (CIPM) \cite{cipm25}.
As the uncertainties of this and other determinations \cite{doe21, lin25a} of the ${}^{171}\mathrm{Yb}^+ (\mathrm{E3}) / {}^{87}\mathrm{Sr}$ frequency ratio are comparable, its inclusion resulted in a substantial shift with respect to the 2021 CIPM evaluation \cite{mar24a}.

More generally, this work connects the two clusters of optical clocks that have been compared with uncertainties below $5 \times 10^{-18}$ to date.
These two clusters are ${}^{27}\mathrm{Al}^+$, ${}^{171}\mathrm{Yb}$ and ${}^{87}\mathrm{Sr}$ clocks \cite{aep26} and ${}^{171}\mathrm{Yb}^+ (\mathrm{E3})$, ${}^{115}\mathrm{In}^+$ and ${}^{88}\mathrm{Sr}^+$ clocks \cite{hau25, fil26}.
Our measurement, thus, creates the opportunity to complete a closure at the level of $5 \times 10^{-18}$ uncertainty, e.g., by measuring the ${}^{27}\mathrm{Al}^+ / {}^{171}\mathrm{Yb}^+ (\mathrm{E3})$ frequency ratio.

In summary, we have reported a high-accuracy optical frequency comparison between the transportable $^{87}$Sr lattice clock, Sr4, and the laboratory $^{171}\mathrm{Yb}^{+} (\mathrm{E3})$ single-ion clock, Yb1E3. 
At a fractional uncertainty of \mbox{$4.3 \times 10^{-18}$}, this measurement is among the most accurate interspecies clock comparisons to date.
The obtained frequency ratio constitutes an accurate benchmark for tests of optical clocks.
Furthermore, we have demonstrated the long-term reproducibility of this frequency ratio. Despite operational breaks, transportation and off-site operation of the lattice clock, the mean values of the individual measurement campaigns differ from the overall average result by no more than $3 \times 10^{-18}$ in fractional units.
Therefore, the transportable clock is well prepared for both chronometric leveling at the centimeter uncertainty level and inter-institute frequency comparisons at the $10^{-18}$ accuracy level, helping to pave the way towards an optical redefinition of the SI second \cite{dim24}.

\textit{Acknowledgments--} We thank Uwe Sterr and Thomas Legero for reliable operation of the ultrastable silicon resonator and Burghard Lipphardt for the operation of an optical frequency comb.
We acknowledge support
by the project 22IEM01 TOCK, which has received funding from the European Partnership on Metrology, co-financed from the European Union’s Horizon Europe Research and Innovation Programme and by the Participating States;
by the German Research Foundation (DFG) under Germany’s Excellence Strategy -- EXC-2123 and EXC-2123/2 QuantumFrontiers -- Project-ID 390837967, SFB~1464 TerraQ -- Project-ID 434617780 -- within project A04 and SFB~1227 DQ-mat -- Project-ID -- within project B02, and by FOR~5456 Project-ID 490990195; and
by the Max Planck–RIKEN–PTB Center for Time, Constants and Fundamental Symmetries.

\section{Supplemental Material}

\setcounter{table}{0}
\renewcommand{\thetable}{A\Roman{table}}
\setcounter{figure}{0}
\renewcommand{\thefigure}{A\arabic{figure}}
\renewcommand{\theequation}{A\arabic{equation}}

\subsection{Determination of the E1 magic frequency in $^{87}$Sr}\label{app:E1MagicFreq}

The detuning of the lattice laser frequency from the E1 magic frequency of the clock is a critical parameter in the evaluation of lattice light shifts in ${}^{87}\mathrm{Sr}$ lattice clocks.
Here, the E1 magic frequency $\nu^{E1}$ is the frequency near 368~THz at which the differential electric-dipole (E1) combined scalar and tensor polarizability of the clock states is zero.
Unlike the scalar polarizability, the tensor polarizability depends on the polarization state of the lattice light and the magnetic quantum number $m_F$ of the atom with respect to the quantization axis set by the bias magnetic field \cite{wes11, shi15}.
In our case, the lattice light is linearly polarized along the bias magnetic field ($\theta = 0^\circ$), and $m_F = m_F^\prime = \pm 9/2$.
If these experimental parameters and the laser frequency are reproduced precisely, we can simplify the evaluation of the lattice light shift \cite{bro17} considerably as light shift measurements can be reused across campaigns.
This may even remove the need for dedicated light shift measurements after transportation, e.g., in a campaign at a remote location.

During each measurement campaign, we have performed interleaved self-comparisons of the clock between two lattice trap depths, while the atomic state distribution was determined from sideband spectra \cite{bla09a}.
While we evaluated the lattice light shifts separately for previous clock comparisons \cite{icon24, nos25}, we have now reviewed those light shift measurements, in order to determine at which level the E1 magic frequency is reproduced in Sr4 and derive an improved evaluation of the lattice light shift by combining the results.

{
\begin{figure}[tb]
    \centering
    \includegraphics[width=0.9\linewidth]{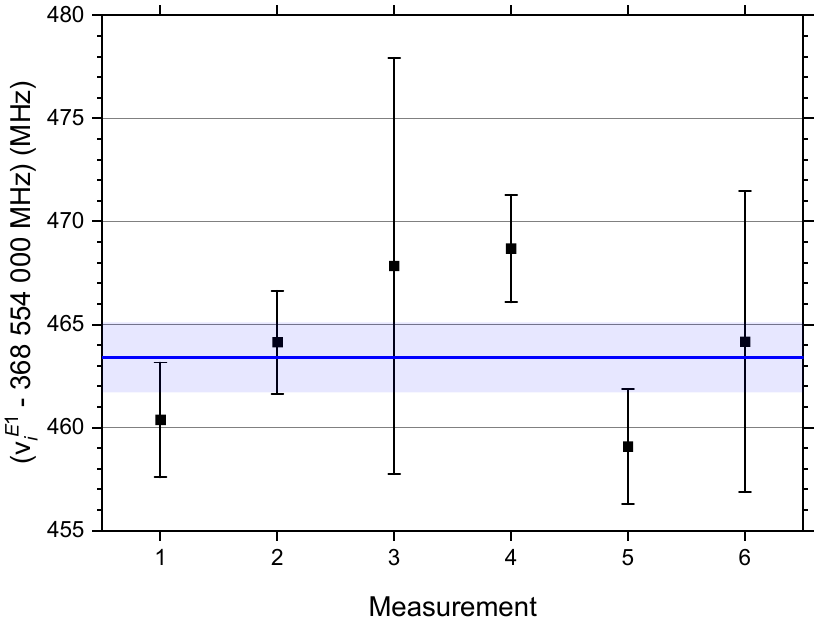}
    \caption{
        Estimated values of the E1 magic frequency of Sr4 per campaign (1 = campaign C1; 3 = campaign C2; 4 = campaign C3).
        The blue solid line and the shaded region correspond to the weighted average and $1\sigma$ confidence interval, respectively.
    }
    \label{fig:E1MagicFrequency}
\end{figure}
}

Such an improved evaluation is only feasible if the absolute frequency of the lattice laser is either reproduced or known, e.g., with respect to a precise reference, for every campaign.
We have thus measured the absolute frequency with respect to an accurate RF reference, provided by hydrogen masers, with an uncertainty of about 100~kHz for every campaign via an optical frequency comb since 2024 (campaigns 1 through 3 and off-site campaigns) to estimate the scalar contribution to the lattice light shift.
We note that the lattice laser's frequency has not been measured during campaign 0, therefore we rely on interleaved measurements from this campaign to determine the lattice light shift as described earlier \cite{icon24, nos25}.

This leaves the reproducibility of the $\theta$-dependent differential tensor polarizability to be investigated.
The polarization of the lattice light is defined by polarizing beam splitters close to the atoms, which enforces linear polarization with well-reproduced orientation.
During operation, we take care to keep stray magnetic fields well compensated. The bias magnetic field is thus given by the coil geometry.
There may be a small offset of $\theta$ from $0^\circ$, away from the maximally insensitive collinear configuration, but it is expected to be reproduced well across campaigns.
We cancel stray magnetic fields by minimizing the spectroscopic linewidth of the clock transition when all ten $m_F$ states are present without a bias field.
Based on an estimated uncertainty of $3^\circ$ for a possible reproducible offset of $\theta$ and on the observed residual linewidths during the magnetic field cancellation measurements, we estimate that variations of the magnetic field orientation cause $\nu^{E1}$ to vary with a standard deviation of $0.28 \, \mathrm{MHz}$ from campaign to campaign.

Figure~\ref{fig:E1MagicFrequency} shows the measurements of $\nu^{E1}$ in individual campaigns, $\nu^{E1}_i$.
The uncertainties include minor contributions from the frequency measurement of the lattice laser and variations of the magnetic field direction that may arise from imperfect magnetic field cancellation.
The scatter of the measurements is higher than expected, with a reduced chi-squared value $\chi^{2}_\mathrm{red} = 1.6$ and a corresponding $p$-value of $0.16$.
The observed overscatter may simply be a statistical fluctuation.
However, if it reflects an underlying issue, it is unlikely that $\nu^{E1}$ itself varied significantly.
Instead, the apparent variation could more plausibly be attributed to imperfections in the parameter-dependent analysis of the measurement data, possibly attributed to atomic temperature determination \cite{han18b, got25}.
To account for this possibility, we conservatively multiply the uncertainties shown in Fig.~\ref{fig:E1MagicFrequency} by $\sqrt{\chi^{2}_\mathrm{red}}$ in the following.

We compute an improved estimate of $\nu^{E1}$ by taking a weighted average of the results $\nu^{E1}_i$.
The individual results are weighted with their respective inverse variances.
While, according to the adopted model, most contributions to the uncertainties of the $\nu^{E1}_i$---including the large ones---are uncorrelated between different campaigns, 
some introduce weak correlations between the results.
The E2-M1 polarizability coefficient and possible systematic offsets in the estimate of the population distribution in the lattice are the most relevant effects that fall into the latter category, both contributing uncertainties well below 1~MHz.
Taking these sources of correlations into account, we find $\nu^{E1} = 368 \, 554 \, 463.4(1.7) \, \mathrm{MHz}$.
This value is in agreement with the one from Ref. \cite{ush18}, which uses the same $m_F$ levels, nominal magnetic field and lattice polarization geometry.
The values reported in Refs. \cite{kim23, jia26b} differ because other $m_F$ levels were addressed.
To reduce the sensitivity of the lattice light shift to fluctuations in the lattice laser intensity \cite{ush18}, the lattice laser had a small positive detuning in the range $+2.1 \ldots +3.4 \, \mathrm{MHz}$ with respect to $\nu^{E1}$ during all measurements described in this manuscript.

This estimate allows us to calculate the lattice light shift in Sr4 using the known detuning from $\nu^{E1}$ in each campaign.
The resulting fractional lattice light shift uncertainties depend on the specific experimental conditions, ranging from \mbox{$1.3 \times 10^{-18}$} to \mbox{$2.8 \times 10^{-18}$}.
For the measurements described here, the lattice light shifts differ by no more than $4.5 \times 10^{-18}$ between this approach and the per-campaign approach \cite{nos25}, which is at the level of the lattice light shift uncertainties.

While updating the lattice light shift evaluation we also update the E2-M1 polarizability coefficient to include recent results \cite{jia26b} and now use $\alpha_\mathrm{qm}/h = -1.064(69) \, \mathrm{mHz}$, which is the weighted average of Refs.~\cite{ush18, kim23, jia26b}. This only marginally changes the value compared to the one used in Ref. \cite{nos25}, corresponding to the weighted average of Refs.~\cite{ush18, kim23}, but it roughly halves its uncertainty.

\setcounter{table}{0}
\renewcommand{\thetable}{B\Roman{table}}
\setcounter{figure}{0}
\renewcommand{\thefigure}{B\arabic{figure}}
\renewcommand{\theequation}{B\arabic{equation}}

\subsection{Correlations concerning the average frequency ratio}\label{app:LocalRatio}

Table \ref{tab:avg-ratio-correlations} reports the correlation coefficients between the average frequency ratio $\overline{R}$ and different quantities.

\begin{table}[ht]
    \begin{center}\small
	\begin{tabular}{p{1in}D{.}{.}{2}}
		\hline
        \hline
		\textbf{Quantity $q$} & \multicolumn{1}{l}{\textbf{Correlation coefficient $r(\overline{R}, q)$}} \rule{0pt}{2.5ex} \\
		\hline
		$\Delta_{B, \mathrm{Yb1E3}}$ &  0.621 \\
		$\Delta_{B, \mathrm{Sr4}}$   & -0.605 \\
		$\Delta_{\mathrm{grav}}$              &  0.115 \\
		$R_0$                                 &  0.649 \\
		$R_1$                                 &  0.801 \\
		$R_2$                                 &  0.750 \\
		$R_3$                                 &  0.943 \\
		\hline
	\end{tabular}
    \end{center}
	\caption{
        Correlation coefficients $r$ of the overall average frequency ratio $\overline{R}$ with respect to the systematic errors ($\Delta_{B,k}$, $k=\mathrm{Yb1E3},\mathrm{Sr4}$), the relativistic redshift error ($\Delta_\mathrm{grav}$), and the per-campaign results $R_i$.
    }
	\label{tab:avg-ratio-correlations}
\end{table}

\end{document}